\documentclass[lettersize,10 pt,conference]{ieeeconf}

\usepackage{bm}

\usepackage{graphicx} 
\usepackage{booktabs} 
\usepackage{caption} 
\usepackage{arydshln} 
\usepackage{float} 
\usepackage{hyperref} 

\usepackage{array}
\usepackage{cite}
\usepackage{amsmath,amssymb,amsfonts}
\usepackage{algorithmic}
\usepackage{graphicx}
\usepackage{threeparttable}
\usepackage{textcomp}
\usepackage{algorithmic}
\usepackage{algorithm}
\usepackage{multirow}
\usepackage{url}
\usepackage{siunitx}
\usepackage{comment}
\usepackage{tikz}
\usepackage{pgf-pie}
\usepackage{pgfplots}
\usepackage{pgfplotstable}
\definecolor{color1}{RGB}{113, 191, 234} 
\definecolor{color2}{RGB}{36, 123, 160} 

\definecolor{darkgreen}{rgb}{0.0, 0.5, 0.0}

\usepackage{caption} 
\usepackage{booktabs} 
\usepackage{xcolor} 

\begin{document}

\title{\LARGE \bf Fault-Tolerant Control for System Availability and Continuous Operation in Heavy-Duty Wheeled Mobile Robots}

\author{Mehdi Heydari Shahna, Pauli Mustalahti, Jouni Mattila
    \thanks{This work was supported by the Business Finland Partnership Project, `Future All-Electric Rough Terrain Autonomous Mobile Manipulators' under Grant No. 2334/31/2022. (Corresponding author: Mehdi Heydari Shahna.)}
\thanks{The authors are with the Faculty of Engineering and Natural Sciences, Tampere University, Tampere, Finland
(e-mail: mehdi.heydarishahna@tuni.fi; pauli.mustalahti@tuni.fi; jouni.mattila@tuni.fi).}%
}

\maketitle

\begin{abstract}
When the control system in a heavy-duty wheeled mobile robot (HD-WMR) malfunctions, deviations from ideal motion occur, significantly heightening the risks of off-road instability and costly damage. To meet the demands for safety, reliability, and controllability in HD-WMRs, the control system must tolerate faults to a certain extent, ensuring continuous operation. 
To this end, this paper introduces a model-free hierarchical control with fault accommodation (MFHCA) framework designed to address sensor and actuator faults in hydraulically powered HD-WMRs with independently controlled wheels. 
To begin, a novel mathematical representation of the motion dynamics of HD-WMRs, incorporating both sensor and actuator fault modes, is investigated. Subsequently, the MFHCA framework is proposed to manage all wheels under various fault modes, ensuring that each wheel tracks the reference driving velocities and steering angles, which are inverse kinematically mapped from the angular and linear velocities commanded in the HD-WMR's base frame. To do so, this framework generates appropriate power efforts in independently valve-regulated wheels to accommodate the adaptively isolated faults, thereby ensuring exponential stability.
The experimental analysis of a 6,500-kg hydraulic-powered HD-WMR under various fault modes and rough terrains demonstrates the validity of the MFHCA framework.
\end{abstract}

\section{Introduction}
Both light-duty and heavy-duty wheeled mobile robots (HD-WMRs) can to perform tasks that are inefficient, unsafe, or unfeasible for humans. However, while light-duty WMRs are typically used in structured environments, like warehouses and laboratories, HD-WMRs excel in challenging applications in rough terrains and under heavy-load conditions \cite{heydari2024radsadobust}. Many HD-WMRs are increasingly equipped with in-wheel hydraulic drive systems, each actuated by its motor and paired with front and rear steering mechanisms to improve maneuverability under heavy loads, enhance responsiveness to off-road conditions, and enable independent power delivery \cite{liikanen2019path}.

However, HD-WMRs face stricter safety and reliability standards than other industrial systems, as operators are often exposed to greater risks due to the absence of full isolation from hazardous conditions \cite{ding2020active}. The primary challenges in this context stem from the high risk of various failures of the operational system \cite{ding2020sensor}. When any stage of the wheel mechanism fails, the robot loses its stability and control while in motion, leading to potential damage \cite{nonami2014hydraulically}. For instance,
sensor faults may arise unexpectedly during the execution of closed-loop control processes for numerous reasons \cite{kelley2020new}. Such failures can compromise system stability and potentially lead to serious accidents because the performance of the control system can be severely degraded, particularly when the fault affects critical sensors essential to its functionality \cite{nahian2020unknown}. Hence, \cite{djordjevic2022sensor} investigated a mechanism for sensor fault estimation by transforming the hydraulic servo model into a new coordinate system to facilitate the observation of sensor faults.
Meanwhile, to mitigate sensor bias and noise under ever-changing driving conditions, \cite{ding2020longitudinal} proposed an enabling multi-sensor fusion-based longitudinal vehicle speed estimator.
In addition, the power effort generated by the actuator module of an independent wheel, which is regulated by valve control signals, is susceptible to faults, either in the hydraulic motor mechanism or due to valve displacement errors, and such faults can compromise the robot's stability, posing a risk of severe accidents \cite{lee2020fault, ding2020active, nonami2014hydraulically}. 

Due to the increasing complexity of hydraulic-powered HD-WMRs, fault handling of such autonomous systems has become a pressing priority to enable control and to improve system dependability.
It plays a vital role in ensuring that systems and equipment operate without failure throughout their life cycles \cite{amiri2023resilient}.
Within the control system community, researchers have concentrated on a particular control design approach known as fault-tolerant control (FTC), which is mainly divided into two categories: passive FTC and active FTC. Active FTC maintains stability and satisfactory performance following a fault by dynamically adjusting the controller through an online fault detection and diagnosis system that identifies and evaluates the fault \cite{jain2018active}. Unlike the active approach, passive FTC relies on a robust controller designed to handle all anticipated faults, integrating real-time fault detection and isolation with control adjustments, allowing the system to maintain acceptable performance despite the occurrence of faults \cite{ji2003hybrid}. Although passive FTC is effective only for faults considered during the design phase, it avoids the time delay associated with fault detection, diagnosis, and controller reconfiguration in active FTC, which is crucial in scenarios where the system can only remain stable for a short period after a fault occurs. In practice, passive FTC often complements active FTC by maintaining system stability during the fault detection and estimation process before transitioning to active FTC for further reconfiguration \cite{benosman2011passive}.

To address the mentioned challenges, this paper introduces a novel model-free hierarchical control with fault accommodation (MFHCA) framework for highly complex and multi-stage HD-WMRs. It aims to accommodate both sensor and actuator faults passively in hydraulically powered
HD-WMRs with independently controlled wheels, while ensuring continuous operation. Contributions of this work are summarized, as follows: 1) a new mathematical representation of the motion dynamics of HD-WMRs, incorporating various sensor and actuator modes, such as healthy, stuck failure, no signal, inefficient, and noise- or disturbance-affected modes, is introduced. This representation lays
the groundwork for designing an effective FTC;
2) although few studies have focused on designing FTC strategies for both sensor and actuator faults in other applications (e.g., \cite{yu2021fractional, meng2020adaptive, zhang2020adaptive, mazare2021fault, wang2020fault}), most existing FTC schemes addressed only sensor faults \cite{baimukashev2021end, kelley2020new, ding2020sensor} or actuator faults \cite{su2023fault, shahna2020anti, bustan2013adaptive, zhao2023adaptive}. However, the MFHCA framework is the first FTC method designed specifically for hydraulic-powered HD-WMR systems, capable of handling both actuator and sensor faults; it features lower implementation complexity and requires only four design parameters for tuning; 3) the MFHCA framework remains entirely independent of any information from the modeling system, and it achieves strong control stability with exponential convergence rate.

The rest of the paper is organized as follows: after a brief explanation of the mechanisms of the studied hydraulic-powered HD-WMR, Section II presents the inverse kinematics of HD-WMRs to map the commanded motion in the robot's base frame to the four driving speeds and rear/front steering angles of independent wheels.
Through an in-depth exploration of the modeling of different sensor and actuator faulty modes, new representations of the motion dynamics of HD-WMRs are given in Section III.
The step-by-step MFHCA framework design and a detailed stability analysis are given in Section IV.
Section V presents experimental fault scenarios exposed to a 6,500-kg HD-WMR under rough and snowy terrains to evaluate the robustness and responsiveness of the MFHCA framework. In this paper, $\mathbb{R}$ is used to denote the set of real numbers, while $\mathbb{R}^n$ refers to an $n$-dimensional real vector space. The symbol $(\cdot)^\top$ represents the transpose operation, and $\lambda_{\max}(\cdot)$ and $\lambda_{\min}(\cdot)$ correspond to the largest and smallest eigenvalues, respectively. The notation $\mathbb{R}^\ell \rightarrow \mathbb{R}^{n}$ denotes a function that is defined from a $\ell$-dimensional real space $\left(\mathbb{R}^\ell\right)$ to a $n$-dimensional real space $\left(\mathbb{R}^n\right)$.
Define ${ }^b \boldsymbol{x}_j^w$ as the variable $\bm{x}$ of the frame $j$ relative to the frame $b$, in the frame $w$.

\section{Inverse Kinematics}
\label{kinema}
\begin{figure}[h!]
\hspace*{-0.0cm} 
\centering
\scalebox{1}{\includegraphics[trim={0cm 0.0cm 0.0cm 0cm},clip,width=\columnwidth]{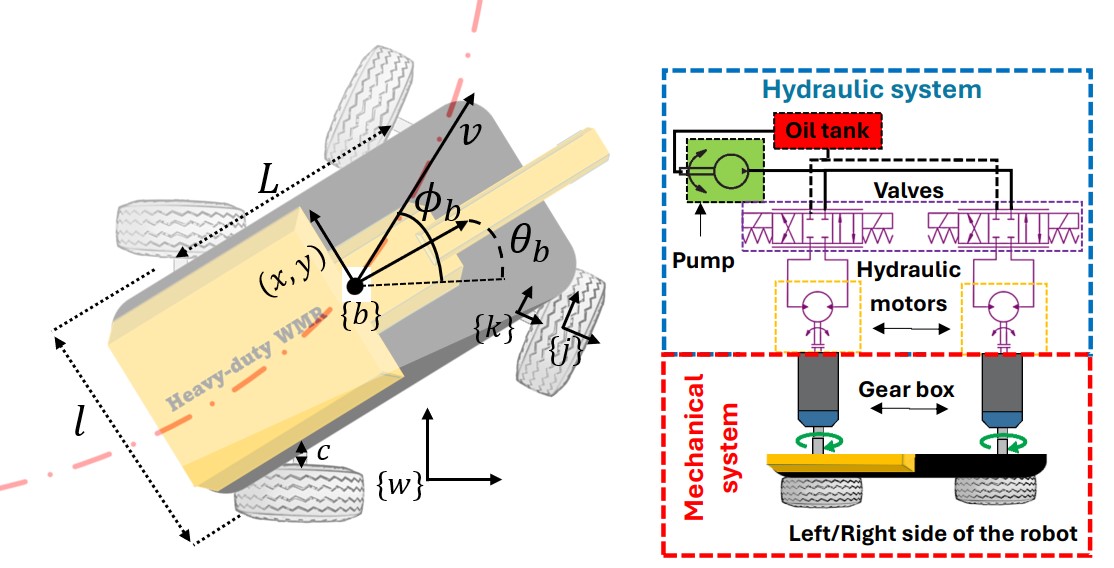}}
\caption{HD-WMR motion with Ackermann steering principle and the mechanism of the drive system.}
\label{kinematic}
\end{figure}

The robot is equipped with four wheels, each independently actuated by fixed-displacement hydraulic motors. The speeds of the motors are adjusted by different servo valves, which translate these adjustments into the corresponding wheel speeds through gear ratios (see the right side of Fig. \ref{kinematic}).
In addition, two identical hydraulic cylinders control the steering of the rear and front wheels, each of which controlled by a servo valve. These cylinders are linked to steering mechanisms that simultaneously move both wheels on the same axis, designed based on the Ackermann steering principle (see the left side of Fig. \ref{kinematic}). The angular and linear velocities of the robot's base frame $b$ are transformed into the wheel velocities and steering angles through inverse kinematics. Fig. \ref{kinematic} depicts the configuration of the robot, which has offset wheels separated by a defined distance from the steering joint $k$ to the wheel center $j$. 

According to \cite{kelly2013mobile, liikanen2019path}, the velocities of frame $\mathrm{b}$ in the world coordinate system $\mathrm{w}$ can be converted to those of frame $\mathrm{j}$ within $\mathrm{w}$, facilitating the calculation of the velocity command of the $i$th wheel, $v_{c,i}$, as shown below:
\begin{equation}
\small
\begin{aligned}
\label{1}
&{ }^b \boldsymbol{v}_j^w={ }^b \boldsymbol{v}_b^w+\boldsymbol{\omega}_b^w \hat{k} \times \boldsymbol{r}_k^b+\boldsymbol{\omega}_j^w \hat{k} \times R_k^b \boldsymbol{r}_j^k \\
&v_{c, i}=\operatorname{sgn}\left(v_c\right) \sqrt{v_{j y}^{w^2}+v_{j x}^{w^2}}
\end{aligned}
\end{equation}
where ${ }^b \boldsymbol{v}_j^w :\mathbb{R}^+ \rightarrow \mathbb{R}$ is the linear velocity of the wheel frame ${\mathrm{j}}$ in the world frame ${\mathrm{w}}$, expressed relative to the base frame ${\mathrm{b}}$. ${ }^b \boldsymbol{v}_b^w :\mathbb{R}^+ \rightarrow \mathbb{R}$ is the velocity of the base frame ${\mathrm{b}}$ in the world frame ${\mathrm{w}}$; $\boldsymbol{\omega}_b^w :\mathbb{R}^+ \rightarrow \mathbb{R}$ is the angular velocity of the base frame ${\mathrm{b}}$ in the world frame ${\mathrm{w}}$; $\hat{k}$ is a unit vector along the rotation axis, vertical for steering; $\boldsymbol{r}_k^b$ is a relative position vector from the base frame ${\mathrm{b}}$ origin to the steering joint ${\mathrm{k}}$;  $\boldsymbol{\omega}_j^w$ is
the angular velocity of the wheel frame ${\mathrm{j}}$ in the world frame ${\mathrm{w}}$, $R_k^b$ is a rotation matrix that transforms coordinates from the steering joint frame ${\mathrm{k}}$ to the base frame ${\mathrm{b}}$; and $\boldsymbol{r}_j^k$ is a relative position vector from the steering joint frame ${\mathrm{k}}$ origin to the wheel center ${\mathrm{j}}$.
In addition, $v_c :\mathbb{R}^+ \rightarrow \mathbb{R} $ represents the size of $v$, $\operatorname{sgn}(v_c)$ is the sign function that determines the direction of the velocity, and $v_{j x}^w, v_{j y}^w$ are the $x$ and $y$ components of the wheel frame ${\mathrm{j}}$ velocity in the world frame ${\mathrm{w}}$.
The equation combines linear and angular velocities, as well as the geometric relationships between frames.
The steering angle command for the $i$th wheel, $\phi_{c,i}:\mathbb{R}^+ \rightarrow \mathbb{R}$, relative to the body frame, is determined based on the velocity of the frame ${\mathrm{k}}$, as:
\begin{equation}
\small
\begin{aligned}
\label{2}
& { }^b \boldsymbol{v}_k^w={ }^b \boldsymbol{v}_b^w+\boldsymbol{\omega}_b^w \hat{k} \times \boldsymbol{r}_k^b, \\
& \phi_{c,i}=\operatorname{atan} 2\left(v_{k y}^w, \quad v_{k x}^w\right) .
\end{aligned}
\end{equation}

\section{Fault-Affected Motion Dynamics}
As the scenarios discussed in this section apply equally to both the steering and driving mechanisms, we will only focus on the driving system to avoid redundancy.

\subsection{Motion Dynamic Model}
\label{1}
Let $i \in\{1, \ldots, 4\}$ denote the order of the wheels, corresponding to the front-right (FR), front-left (FL), rear-right (RR), and rear-left (RL) wheels, respectively.
The motion dynamics of the driving system for each wheel can then be described, as follows:
\begin{equation}
\small
\begin{aligned}
\label{3}
&J_i
\dot{\omega}_i
=
{u}_{v,i}
- {C_m}_i
\omega_i
- {f}_i
- {G}_i
- {{T}_L}_i
\end{aligned}
\end{equation}
where $\omega_i :\mathbb{R}^+ \rightarrow \mathbb{R}$ is the angular velocity of the wheel. $J_i:\mathbb{R}^+ \rightarrow \mathbb{R}$ characterizes the mass (inertia) properties, while ${{C_m}_i} :\mathbb{R}^+ \times \mathbb{R} \rightarrow \mathbb{R}$ accounts for the centrifugal and Coriolis forces. ${G_i}:\mathbb{R}^{+} \times \mathbb{R} \rightarrow \mathbb{R}$ represents the gravitational forces, and ${f_i}:\mathbb{R}^{+} \times \mathbb{R} \rightarrow \mathbb{R}$ accounts for the resistance encountered during movement. $u_{v,i}:\mathbb{R}^{+} \times \mathbb{R} \rightarrow \mathbb{R}$ represents the driving valve control signal applied at the $i$th independent wheel, and ${{{T}_L}_i}:\mathbb{R}^+ \rightarrow \mathbb{R}$ signifies external disturbances. 

\subsection{Sensor Faults}
\label{sensor}
During sensor faults, $\omega_i$, the angular velocity received from the speed sensor, does not represent the actual velocity of the wheel. Thus, we mathematically describe $\omega_i$ as \cite{shahna2024exponential, shahna2021design}:
\begin{equation}
\small
\begin{aligned}
\label{4}
&{\omega_i} = (1-{{{\epsilon}}_{s,i}})\omega_{a,i} + {{{\epsilon}}_{s,i}}\omega_{{\text{sat}},i}\\
\end{aligned}
\end{equation}
where ${{\omega}_{a,i}}:\mathbb{R}^+ \rightarrow \mathbb{R}$ represents the actual velocity of the wheel. $0\leq{{{\epsilon}}_i}_s\leq1$ and ${{\omega}_i}_{\text{sat}}:\mathbb{R}^+ \rightarrow \mathbb{R}$ characterize various types of sensor failures, as shown in Table \ref{sens}.

\begin{table}[h!]
\caption{Descriptive characteristics in sensor faults}
\centering
\small
\begin{tabular}{|c|c|}
\toprule
\textbf{Mathematical condition} & \textbf{Sensor status} \\
\midrule
$\epsilon_{s,i} = 0$ & {Accurate} \\ 
$\omega_{\text{sat},i} \neq 0$ and $\epsilon_{s,i} = 1$ & \textcolor{red}{Stuck failure} \\ 
$\omega_{\text{sat},i} = 0$ and $\epsilon_{s,i} = 1$ & \textcolor{red}{No signal (access)} \\ 
$0 < \epsilon_{s,i} < 1$ and $\omega_{\text{sat},i} = 0$ & {Inefficient} \\
$0 < \epsilon_{s,i} < 1$ and $\omega_{\text{sat},i} \neq 0$ & {Noise-affected}\\
\bottomrule
\end{tabular}
\label{sens}
\end{table}

\textit{Assumption. 1.} Two sensor fault occurrences,`Stuck failure' and `No signal', deliver no closed-loop information from the affected sensor. For these types of faults, if the fault duration is brief, the system may continue functioning without significant issues. However, if the faults persist, operations must be halted, and mechanical repairs are required.

Assume all wheels have the same diameter $r$. From \eqref{4}, we have the fault effects in the linear velocity $v_i = r \omega_i$, as:

\begin{equation}
\small
\begin{aligned}
\label{5}
v_i= \delta_i v_{a,i} + \bar{\delta}_i 
\end{aligned}
\end{equation}
where $\delta_i = (1 - \epsilon_{c,i})$ and $\bar{\delta}_i = r \epsilon_{c,i} \omega_{\text{sat},i}$ are unknown and non-constant. Now, from \eqref{3}:
\begin{equation}
\small
\begin{aligned}
\label{7}
J_i \hspace{0.1cm} \hspace{0.1cm} \delta_i \hspace{0.1cm}
{\dot{v}}_{a,i}
=&
r{u}_{v,i}
- {C_m}_i {v}_{i} - r{f}_i- r{G}_i- r{{T}_L}_i - J_i \frac{\dot{\bar{\delta}}_i}{r}
\\
& -J_i \dot{\delta}_i v_{a,i}
\end{aligned}
\end{equation}

\subsection{Actuator Faults}
During actuator faults, the valve control signal $u_{v,i}$, does not represent the actual signal conveyed to the dynamic system. Thus, assume that ${{u}_{a,i}}:\mathbb{R}^{+} \times \mathbb{R} \rightarrow \mathbb{R}$ represents the control signal commanded by the control system. The actual control signal $u_{v,i}$ is mathematically described, as follows \cite{shahna2020anti, shahna2021design}:
\begin{equation}
\small
\begin{aligned}
\label{8}
u_{v,i} = (1-\epsilon_{c,i})u_{a,i} + \epsilon_{c,i}{{u}_{\text{sat},i}}
\end{aligned}
\end{equation}
where $0\leq{\epsilon}_{i,c}\leq 1$ and ${{u}_{\text{sat},i}}:\mathbb{R}^+ \rightarrow \mathbb{R}$ are unknown variables to characterize various types of actuator failures (see Table \ref{act_conditions}).

\begin{table}[h!]
\caption{Descriptive characteristics in actuator faults}
\centering
\small
\begin{tabular}{|c|c|}
\toprule
\textbf{Mathematical condition} & \textbf{Actuator status} \\ \midrule
$\epsilon_{c,i} = 0$ & Health \\ 
$u_{\text{sat},i} \neq 0$ and ${\epsilon_{c,i}} = 1$ & \textcolor{red}{Stuck failure}\\ 
${u_{\text{sat},i}} = 0$ and ${\epsilon_{c,i}} = 1$ & \textcolor{red}{No control signal} \\ 
$0 < {\epsilon_{c,i}} < 1$ and ${u_{\text{sat},i}} = 0$ & Inefficient \\ 
$0 < {\epsilon_{c,i}} < 1$ and ${u_{\text{sat},i}} \neq 0$ & Disturbance-affected \\ \bottomrule
\end{tabular}
\label{act_conditions}
\end{table}

By inserting the actuator fault model \eqref{8} into the dynamic model system affected by the sensor faults \eqref{7}, we obtain:
\begin{equation}
\small
\begin{aligned}
\label{9}
J_i \delta_i  \hspace{0.1cm} \dot{v}_{a,i}
=&
r(1-\epsilon_{c,i})u_{a,i} + r\epsilon_{c,i}{{u}_{\text{sat},i}}
- {C_m}_i {v}_{i} - r{f}_i- r{G}_i
\\
&- r{{T}_L}_i - J_i \frac{\dot{\bar{\delta}}_i}{r}-J_i \dot{\delta}_i v_{a,i}
\end{aligned}
\end{equation}

Now, we have the new dynamic model of the
wheel, considering both sensor and actuator fault models, as:
\begin{equation}
\small
\begin{aligned}
\label{10}
 \dot{v}_{a,i}
=&
\frac{r(1-\epsilon_{c,i})}{J_i \delta_i} u_{a,i} 
- (J_i \delta_i)^{-1} {C_m}_i {v}_{i} -  \delta^{-1}_i \frac{\dot{\bar{\delta}}_i}{r}\\
&-\delta^{-1}_i \dot{\delta}_i v_{a,i}+ r (J_i \delta_i)^{-1} [\epsilon_{c,i}{{u}_{\text{sat},i}}-{f}_i- {G}_i- {{T}_L}_i] 
\end{aligned}
\end{equation}

\textit{Remark. 1.} Without sensor and actuator faults, the dynamic representation provided in \eqref{10} is equivalent to the standard motion dynamics in \eqref{3}.

\textit{Assumption. 2.} Two actuator fault occurrences, `Stuck Failure' and `No Control Signal,' transmit no closed-loop control command signal to the affected valve. This implies that the actual valve signal, $u_{v,i}$, does not vary in response to the commanded valve control, $v_{a,i}$. In other words, the control system is no longer operational, and the system cannot continue functioning without significant issues. Operations must be halted immediately, and mechanical repairs are required. 

\section{The MFHCA Framework Design}
\subsection{Driving Valve Control in the MFHCA Framework}
\label{2}
Assume the reference linear velocity for each wheel $v_{c,i}$, generated by the inverse kinematics in Section \ref{kinema}, is differentiable (the reference acceleration is definable). Define the tracking error of the $i$th wheel as ${e}_i = v_{a,i}- v_{c,i}$. By adding $-\dot{v}_{c,i}$ to both sides of \eqref{10}, the equation becomes:
\begin{equation}
\small
\begin{aligned}
\label{13}
 \dot{e}_i
=a_i u_{a,i}+F_i + D_i
\end{aligned}
\end{equation}
where:
\begin{equation}
\small
\begin{aligned}
\label{12}
a_i =& \frac{r(1-\epsilon_{c,i})}{J_i \delta_i}, \hspace{0.1cm} F_i = - (J_i \delta_i)^{-1} {C_m}_i {v}_{i}-\delta^{-1}_i \dot{\delta}_i v_{a,i}\\
D_i =& r (J_i \delta_i)^{-1} \epsilon_{c,i}{{u}_{\text{sat},i}}- r (J_i \delta_i)^{-1} [{f}_i+ {G}_i+ {{T}_L}_i]
\\
& -  \delta^{-1}_i \frac{\dot{\bar{\delta}}_i}{r}- \dot{v}_{c,i}
\end{aligned}
\end{equation}

As we aim to design a model-free control, we assume that $F_i :\mathbb{R}^{+} \times \mathbb{R} \rightarrow \mathbb{R}$ and $D_i :\mathbb{R}^{+} \times \mathbb{R} \rightarrow \mathbb{R}$ are unknown for the MFHCA framework. In addition, the coefficient of valve control $a_i :\mathbb{R}^+ \rightarrow \mathbb{R}^{+}$ is an unknown and non-constant positive variable. 
Let us define a quadratic function for each wheel equipped with an independent hydraulic motor:
\begin{equation}
\small
\begin{aligned}
\label{14}
{V}_i = \hspace{0.1cm} \frac{1}{\bar{a}_i} {e^2_{i}}+ \frac{1}{2} \hat{\Psi}_{i}^2
\end{aligned}
\end{equation} 
where $\bar{a}_i \in \mathbb{R}^{+}$ is the unknown upper bound of $a_i$, and $\hat{\Psi}_i$ is the $i$th adaptive parameters, which is defined, as:
\begin{equation}
\small
\begin{aligned}
\label{16}
\dot{\hat{\Psi}}_i=(-\beta_{i}-|v_{c,i}|- |v_i| - \xi_i)\hat{\Psi}_{i}
\end{aligned}
\end{equation}
where $\xi_i$ and $\beta_i\in \mathbb{R}^{+}$ are positive constants. Let us propose the following driving valve-regulated control for the $i$th wheel:
\begin{equation}
\small
\begin{aligned}
\label{15}
{u}_{a,i} =&  -\lambda_i |v_i|+ \lambda_i |v_{c,i}|+ \hat{\Psi}^2_{i}
\end{aligned}
\end{equation} 
where $\lambda_i\in \mathbb{R}^{+}$ is a positive constant.

\textit{Theorem. 1.} Employing \eqref{15} and \eqref{16} ensures the Euclidean norm of the tracking velocity error vector $\bm{e}=\left[e_1, \ldots, e_4\right]^T$, which includes all four wheels under both sensor and actuator faults, exponentially converges to a stable region. The radius of this region is dependent on the intensity of fault occurrences.

\textit{Proof:} See Appendix A.

\subsection{Steering Valve Control in the MFHCA Framework}
\label{steering_contol}
\begin{figure}[h!]
\hspace*{-0.0cm} 
\centering
\scalebox{1.0}{\includegraphics[trim={0cm 0.0cm 0.0cm 0cm},clip,width=\columnwidth]{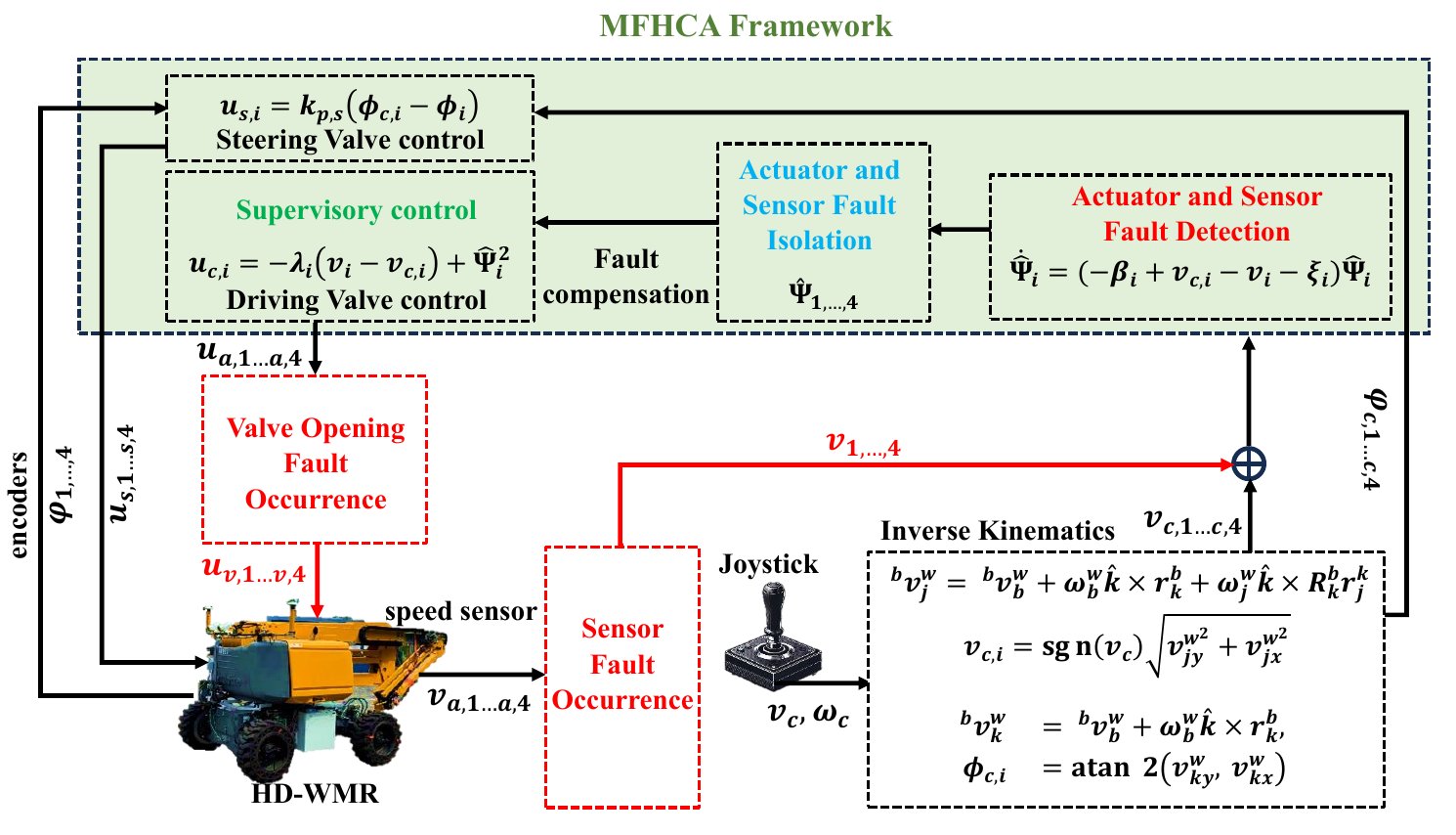}}
\caption{The MFHCA framework-applied HD-WMR system.}
\label{control.s}
\end{figure}

We can also apply the same scenario from Section \ref{1} to Section \ref{2} steering systems. To avoid redundancy, we will omit a discussion of the faults in the steering system and instead use the steering valve control openings provided in \cite{liikanen2019path} within the MFHCA framework. 
Because the steering system is a linear actuation system with lower complexity compared to the driving system, and based on our experience, using an advanced robust control approach may not significantly improve the healthy steering system, so the steering valve control provided in \cite{liikanen2019path} is sufficient. 
The steering valves installed on the axles of the rear and front wheels are managed by adjusting their openings $u_{s, i}:\mathbb{R}^+ \times \mathbb{R}\rightarrow \mathbb{R}$. The actuator control unit processes the steering-angle command $\phi_{c,i}$ generated from Section \ref{kinema}, along with their respective measured signals $\phi_i:\mathbb{R}^+ \rightarrow \mathbb{R}$. Therefore, the following equation for the steering valve control mechanism is provided, as in \cite{liikanen2019path}:
\begin{equation}
\small
\begin{aligned}
\label{17}
u_{s, i}=k_{p, s}\left(\phi_{c,i}-\phi_i\right),
\end{aligned}
\end{equation}
where $k_{p, s}$ is the proportional gain. The MFHCA framework-applied HD-WMR system is shown in Fig. \ref{control.s}.

The sensor and actuator fault scenarios are derived from Eqs. \eqref{4} and \eqref{8}, applied to the input and output of the driving valve control system in the HD-WMR. The linear velocity $v_c$ and angular velocity $\omega_c$ commands from a human-operated joystick are input into the inverse kinematics (Section \ref{kinema}). This section generates the reference velocities $v_{c, i}$ and the corresponding steering-angle commands $\phi_{c, i}$. 
The adaptive law proposed in Eq. \eqref{16} is responsible for detecting and isolating faults, which are then provided to the supervisory control in Eq. \eqref{15} for compensation. Finally, the MFHCA framework generates appropriate power efforts in driving valve control signals $u_{a, i}$ (Eqs. \eqref{15}) and steering valve control signals $u_{s, i}$ (Eq. \eqref{17}) to independent wheels, adaptively accommodating both sensor and actuator fault occurrences.

\section{Experimental Results}

\subsection{Experimental Setup}
\begin{figure}[h!]
\hspace*{-0.0cm} 
\centering
\scalebox{0.8}{\includegraphics[trim={0cm 0.0cm 0.0cm 0cm},clip,width=\columnwidth]{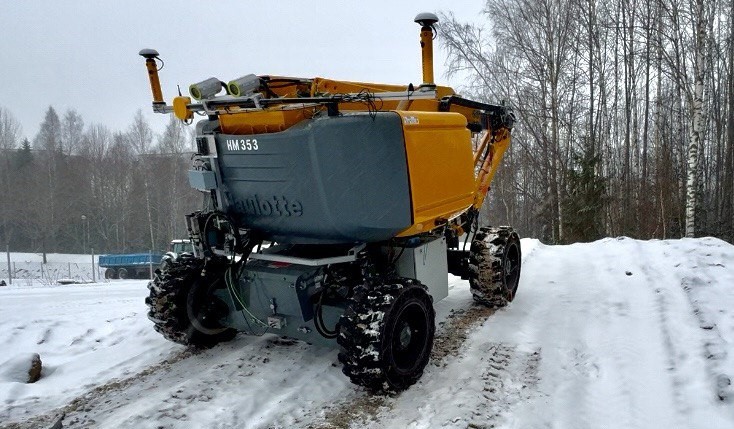}}
\caption{The test-case HD-WMR.}
\label{case}
\end{figure}

In this paper, the Haulotte 16RTJ PRO, a heavy-duty articulated boom lift with a four-wheel-drive robot configuration and a weight of $6,650$ kg, serves as the test-case robot. The image of the test case for HD-WMR is presented in Fig. \ref{case}. Specific information for the instrumentation and hardware of the test-case HD-WMR are summarized in Table \ref{tab:hardware1}.

\begin{table}[h!]
\centering
\caption{Instrumentation and hardware of the HD-WMR}
\small
\begin{tabular}{|c|c|}
\toprule
\textbf{Component} & \textbf{Description} \\
\midrule
Kubota Diesel Engine & $26.5 \mathrm{kW}$ @ $2,800 \mathrm{rpm}$ \\
Bosch Rexroth Pump & $63$ $\mathrm{l} / \mathrm{min}$ \\
Danfoss OMSS Motors & $100$ $\mathrm{cm}^3 / \mathrm{rev}$ \\
Bosch Rexroth valves& $40$ $\mathrm{l} / \mathrm{min} @ \Delta p=3.5 \mathrm{MPa}$ \\
IFM PA3521 transducers& Sensor range: 25 MPa \\
Danfoss EMD Speed Sensor  & $0–2500$ rpm \\
Beckhoff IPC CX2030 & $1,000$-$\mathrm{Hz}$ sample rate \\
\bottomrule
\end{tabular}
\label{tab:hardware1}
\end{table}

It has a wheelbase of $2.1$ m, steering joint spacing of $1.46$ m, wheel diameter of $0.854$ m, a gear ratio of $17.7$, and capabilities of a $0.36$-m/s linear velocity and $45^{\circ}$ steering angle. Its functionality is driven by a variable-displacement hydraulic pump connected to a diesel engine. This pump, controlled by a pressure-regulating servo valve, delivers consistent pressure to ensure reliable functionality, enabling the hydraulic system to execute driving and steering commands effectively. 

\begin{figure*}[h!]
\hspace*{-0.0cm} 
\centering
\scalebox{1.4}{\includegraphics[trim={0cm 0.0cm 0.0cm 0cm},clip,width=\columnwidth]{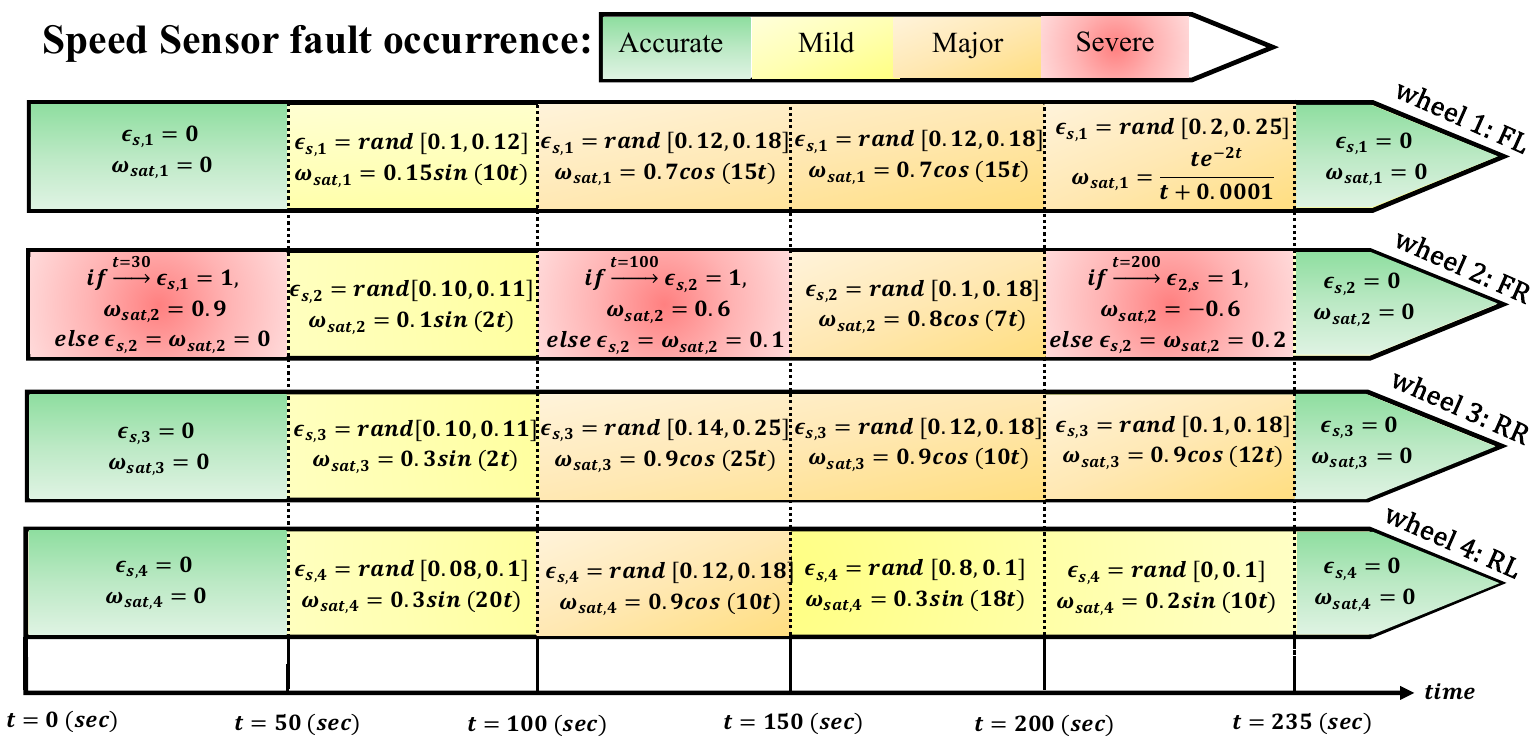}}
\caption{Different fault intervals in four magnet-based speed sensors.}
\label{faulty_s}
\end{figure*}

\begin{figure*}[h!]
\hspace*{-0.0cm} 
\centering
\scalebox{1.400}{\includegraphics[trim={0cm 0.0cm 0.0cm 0cm},clip,width=\columnwidth]{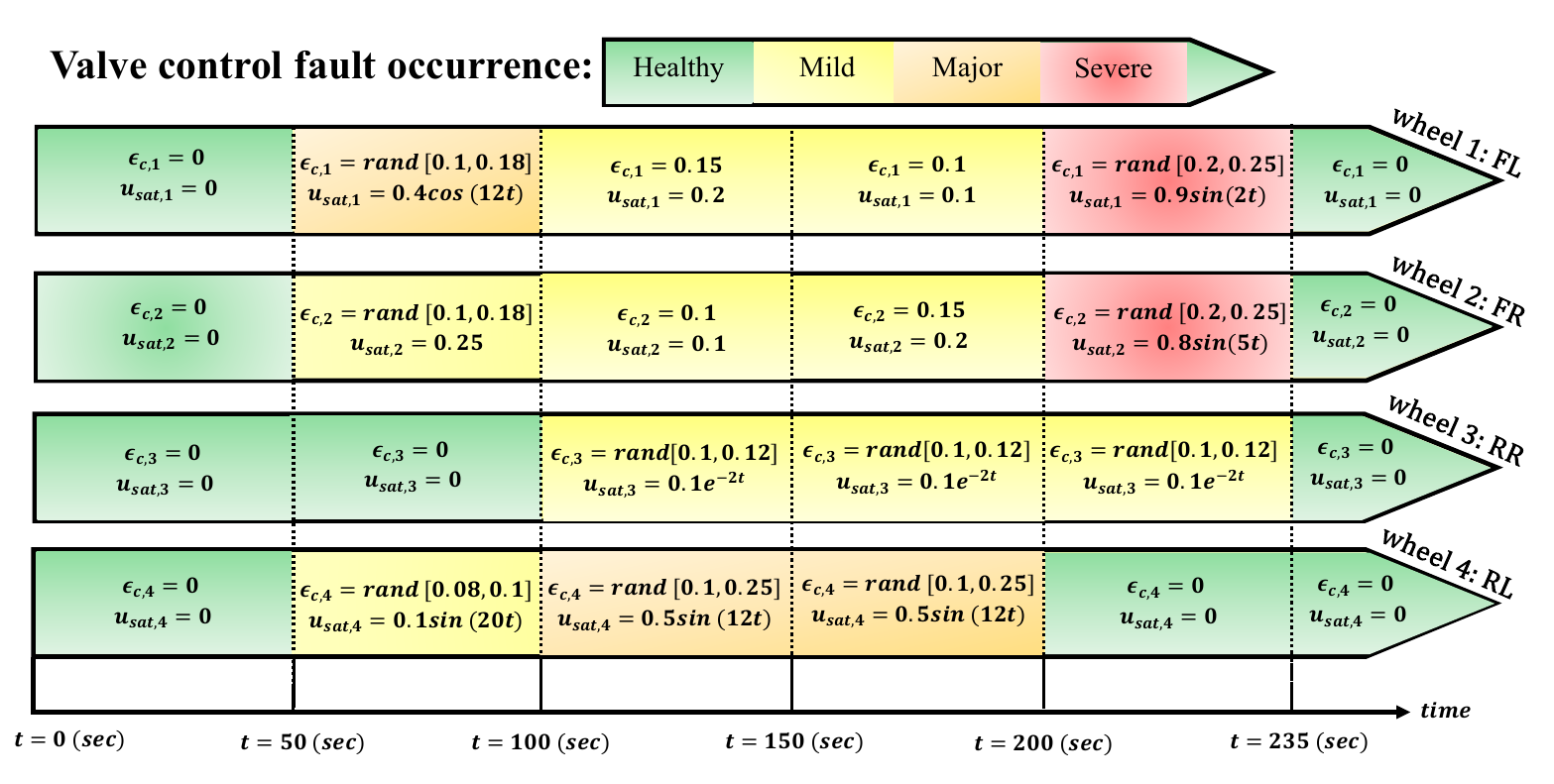}}
\caption{Different fault intervals in four valve-regulated actuators.}
\label{faulty_A}
\end{figure*}

A speed sensor installed on the motor monitors the shaft's rotational speed and direction by using a magnet that spins inside the motor. Two identical hydraulic cylinders, controlled by a servo valve, steer both the front and rear wheels. The cylinders are linked to the steering mechanisms, synchronizing the movement of wheels on the same side. The system follows the Ackermann steering geometry. Wire encoders track the positions of the steering cylinders, and the steering angles are determined using these position measurements combined with the known dimensions of the steering system components.

\subsection{Implementation of the MFHCA Framework }
\begin{figure}[h!]
\hspace*{-0.0cm} 
\centering
\scalebox{0.6}{\includegraphics[trim={0cm 0.0cm 0.0cm 0cm},clip,width=\columnwidth]{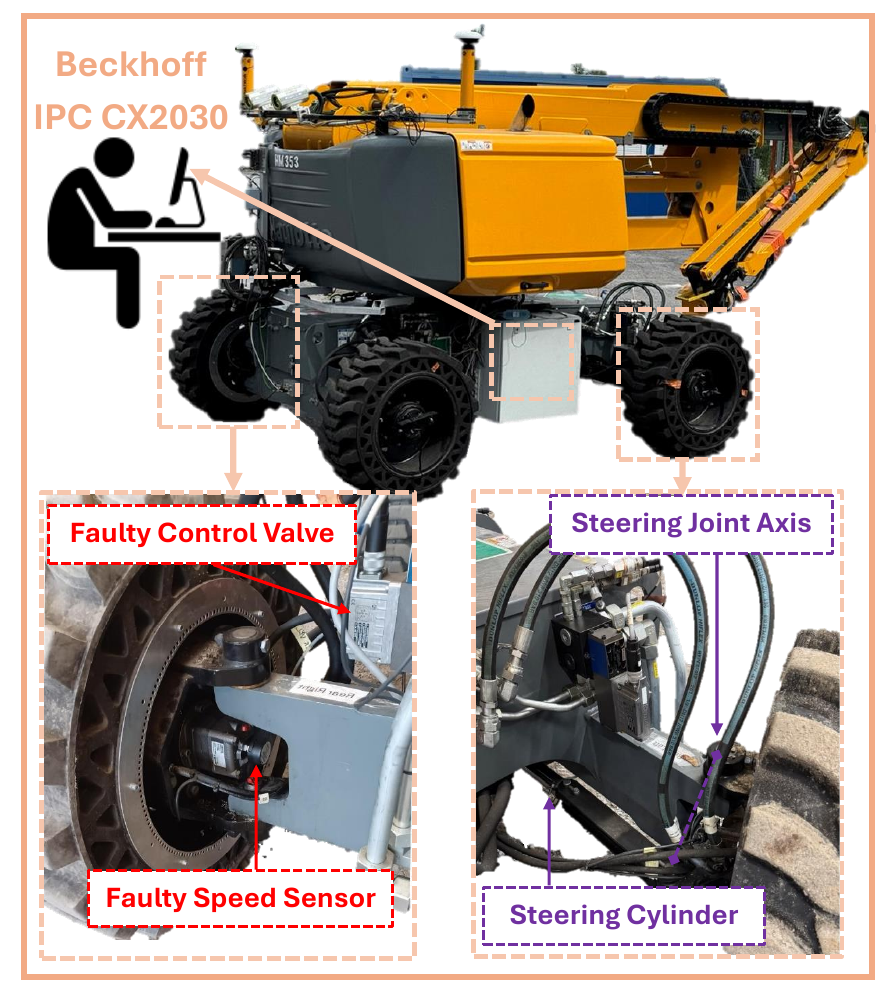}}
\caption{The communication, sensor, and actuator setup.}
\label{faulty_beck}
\end{figure}

\begin{figure}[h!]
\hspace*{-0.0cm} 
\centering
\scalebox{1.0}{\includegraphics[trim={0cm 0.0cm 0.0cm 0cm},clip,width=\columnwidth]{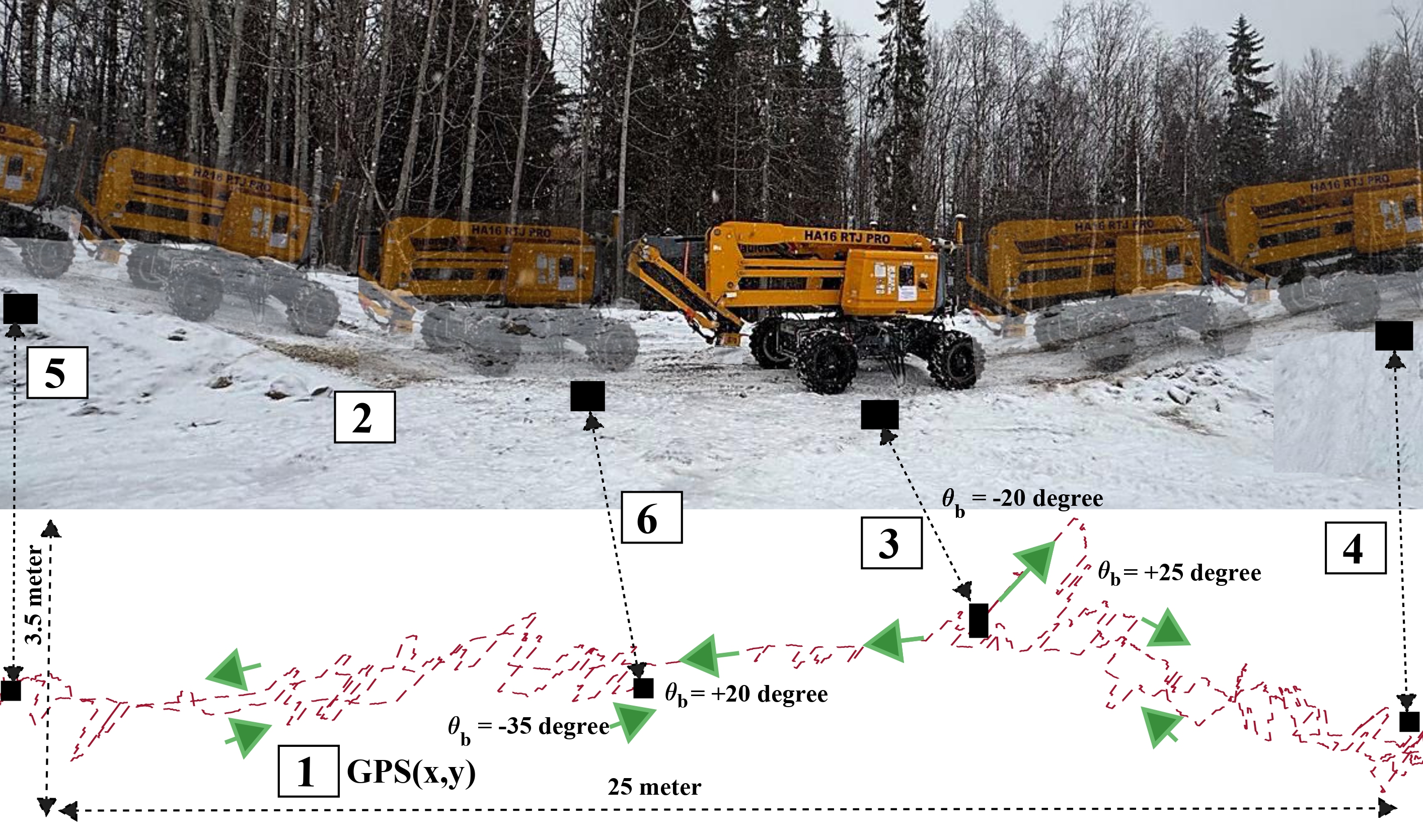}}
\caption{Backward and forward duty cycle: 1) GPS-based path of the HD-WMR; 2) image of the icy and rough terrain at $-8^{\circ} \mathrm{C}$ ; 3) initial position of the robot at $t=0$ seconds; 4) highest point on the right side of the operation at $t=100$ seconds; 5) highest point on the left side of the operation at $t=$ 185 seconds; 6) final position at $t=240$ seconds.}
\label{duty}
\end{figure}

\begin{figure}[h!]
\hspace*{-0.0cm} 
\centering
\scalebox{0.7}{\includegraphics[trim={0cm 0.0cm 0.0cm 0cm},clip,width=\columnwidth]{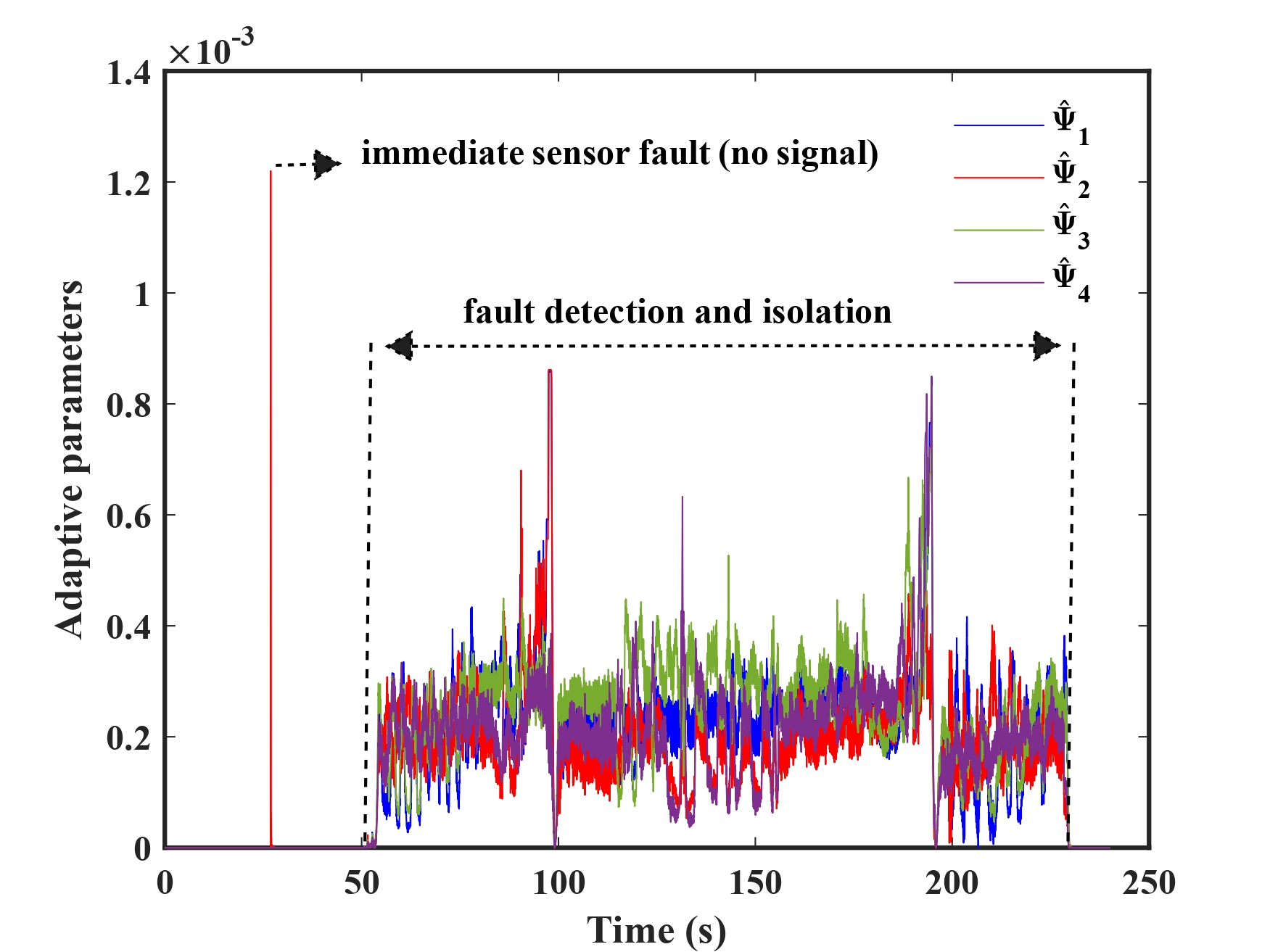}}
\caption{Adaptive law signals provided in Eq. \eqref{16}.}
\label{theta}
\end{figure}

To investigate the robustness and tolerance of the MFHCA framework in the studied HD-WMR, we introduced artificial sensor and actuator fault scenarios for all four wheels at different time intervals, as illustrated in Figs. \ref{faulty_s} and \ref{faulty_A}. The input of the MFHCA framework includes: 1) the errors between the faulty linear velocities $v_{i}$, derived from the faulty magnet-based speed sensor, and the command linear velocities in the wheel frame $v_{c, i}$, being inverse kinematically mapped from the reference velocities $v_c$ and $\omega_c$ in the WMR's base frame; 2) the errors between the steering angles $\phi_{i}$, derived from wire-based encoders, and the reference angular velocities in the wheel frame $\phi_{c, i}$, being inverse kinematically mapped from $v_c$ and $\omega_c$.
Finally, the MFHCA framework generates appropriate power efforts $u_{a, i}$ and $u_{s, i}$ in independent valve-regulated wheels. The parameters of the MFHCA framework used in this study are as follows: $\lambda_i=1, \xi_i=1$, $\beta_i=0.001$, and $k_{p,s}=1$. Communication among components of test-case HD-WMR was established in the Beckhoff IPC CX2030 before the operation (see Fig. \ref{faulty_beck}). The duty cycle of the experiment, conducted under the fault scenarios, is presented in Fig. \ref{duty}. As indicated, the duty cycle was carried out on icy and rough terrain at $-8^{\circ} \mathrm{C}$, involving backward and forward movements, including the ascent and descent of two steep slopes from both the front and rear sides. The maximum commanded rotation of the HD-WMR, resulting from the joystick commands $\phi_{c,i}$, was $\theta_b=35$ degrees.
Fig. \ref{theta} shows the adaptive parameters $\hat{\Psi}_i$ during the duty cycle for detecting and isolating faults. The wheel (FR) experienced a significant jump due to the immediate stuck failures at $t=30$, $100$, and $200$. The results indicate that most short-term compensation efforts were directed at the second wheel, while most long-term efforts focused on the third wheel, due to disturbances caused by ice slippage and the combined effects of sensor and actuator faults.

\begin{figure}[h!]
\hspace*{-0.0cm} 
\centering
\scalebox{0.75}{\includegraphics[trim={0cm 0.0cm 0.0cm 0cm},clip,width=\columnwidth]{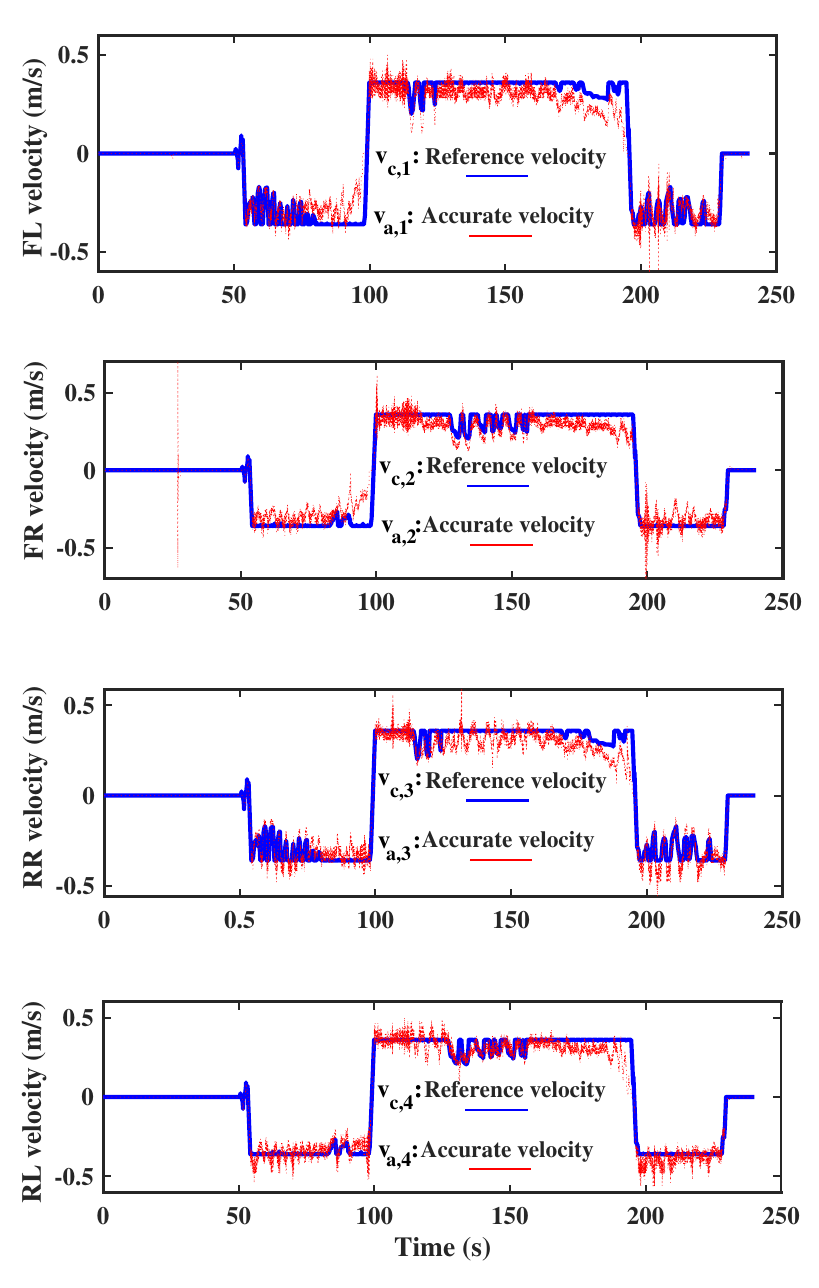}}
\caption{Tracking velocity of the four wheels employing \eqref{15}.}
\label{vel_driv}
\end{figure}

\begin{figure}[h!]
\hspace*{-0.0cm} 
\centering
\scalebox{0.65}{\includegraphics[trim={0cm 0.0cm 0.0cm 0cm},clip,width=\columnwidth]{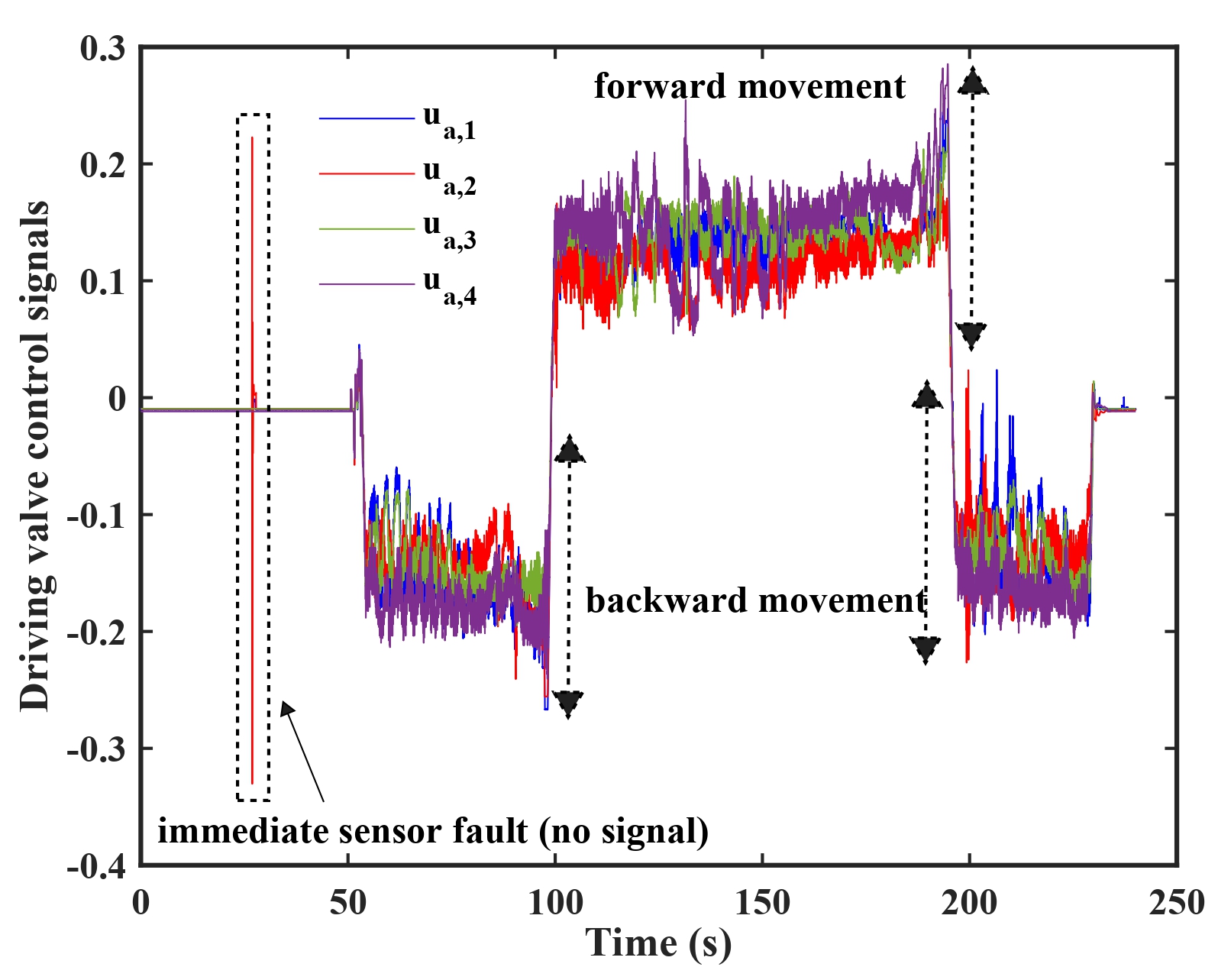}}
\caption{Four driving valve control signals $u_{a, i}$ using Eq. \eqref{15}.}
\label{vavle_drive}
\end{figure}

Fig. \ref{theta} illustrates fault detection between $t=50$ and $t=235$, plus the immediate stuck failure in the sensor of the FR wheel at $t=30$, adhering to the fault scenarios indicated in Figs. \ref{faulty_s} and \ref{faulty_A}.
The tracking of linear velocities for the four wheels is shown in Fig. \ref{vel_driv}, illustrating how the wheels adhered to the commands even under fault conditions. Fig. \ref{vavle_drive} illustrates the four driving valve control signals $u_{a, i}$, derived from the output of Eq. \eqref{15}, to force velocities of wheels $v_{a,i}$ to track the reference linear velocities $v_{c,i}$ while compensating for fault effects. 

\begin{figure}[h!]
\hspace*{-0.0cm} 
\centering
\scalebox{0.65}{\includegraphics[trim={0cm 0.0cm 0.0cm 0cm},clip,width=\columnwidth]{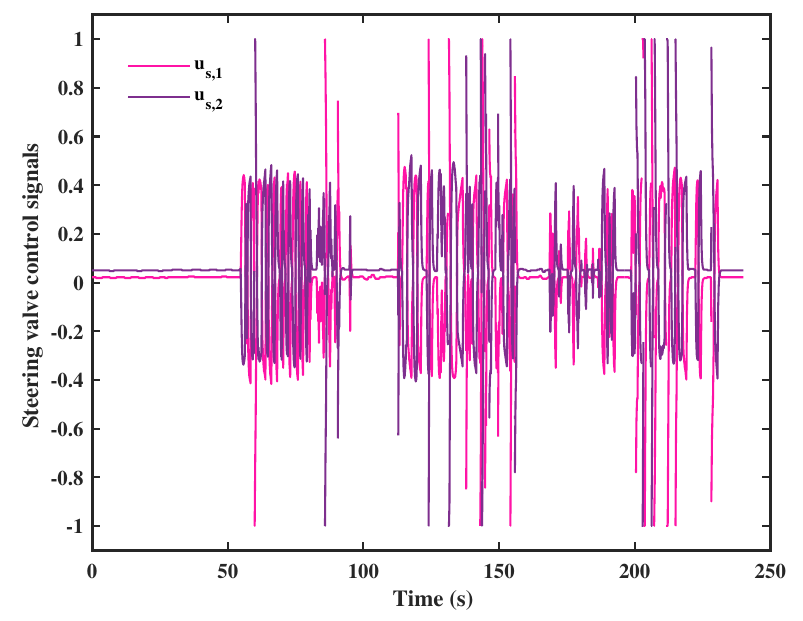}}
\caption{The proposed steering tracking control signals $u_{s,i}$}
\label{steeri_valve}
\end{figure}

The change in the sign of these signals indicates a change in the direction of the HD-WMR and highlights that the majority of efforts occurred during fault periods. In addition, Fig. \ref{steeri_valve} illustrates the steering valve-regulated control proposed in Eq. \eqref{17}. The input is the information of the wire encoders, which show the positions of the two front and rear steering valve cylinders (converted into wheel angles $\phi_{i}$), and the output includes the proposed steering tracking control signals $u_{s,i}$ to track joy-stick-based commands $\phi_{c,i}$.
Table \ref{rrt} presents the comparative performance of state-of-the-art controllers for HD-WMRs, validating the capabilities of the MFHCA framework with fewer design parameters. It demonstrates better average tracking error and reduced valve control effort across all four wheels under identical conditions and snowy terrains. In addition, the implementation complexity is lower due to the reduced number of control design parameters that require tuning. Note: `avg.' refers to average, `no.' to number, `trk. err.' to tracking error, `ctrl. eff.' to control effort, and `des. params' to the number of the required design parameters.     

\begin{table}[h!]
\captionsetup{position=top}
\caption{Comparative performance of state-of-the-art controllers for HD-WMRs in the same condition.}
\centering
\scriptsize
\begin{tabular}{c|c|c|c|c}
\toprule
\toprule
\textcolor{black}{\textbf{No.}} & \textcolor{black}{\textbf{ctrl.}} & \textcolor{black}{\textbf{trk. err. (avg.)}} & \textcolor{black}{\textbf{ctrl. eff. (avg.)}} & \textcolor{black}{\textbf{des. params.}} \\
\midrule
\midrule
\textbf{1} & \textcolor{black}{MFHCA} & {$0.019$ m/s} & {$0.125$} & $4$ \\
\bottomrule
\textbf{2} & \cite{yang2022adaptive} & {$0.032$ m/s} & {$0.130$} & \textcolor{black}{16} \\
\bottomrule
\textbf{3} & \cite{wang2021finite} & {$0.021$ m/s} & {$0.145$} & \textcolor{black}{7} \\
\bottomrule
\textbf{4} & \cite{yu2020adaptive} & {$0.024$ m/s} & {$0.125$} & \textcolor{black}{6} \\
\bottomrule
\textbf{5} & \cite{sun2020fuzzy} & {$0.035$ m/s} & {$0.135$} & \textcolor{black}{12} \\
\bottomrule
\bottomrule
\end{tabular}
\label{rrt}
\end{table}

\section{Conclusion}
To meet the demands for safety, reliability, and controllability in hydraulically powered HD-WMRs with independently controlled wheels, this paper introduced an MFHCA framework with exponential stability. First, a novel mathematical representation of the motion dynamics of HD-WMRs, incorporating various sensor and actuator fault modes, laid the groundwork for designing the MFHCA framework. Then, the MFHCA framework managed all wheels, adaptively accommodating various fault modes, to track the reference velocities and steering angles, which were inverse kinematically mapped from the angular and linear velocities commanded in the HD-WMR's base frame.
The experimental evaluation on a 6,500-kg hydraulic-powered HD-WMR validated the framework’s effectiveness in maintaining stability and controllability across diverse fault scenarios and harsh terrains. This work may pave the way for safer and more efficient operations in challenging environments. Future work may explore the scalability of this framework to other robotic platforms and fault conditions, further advancing the field of fault-tolerant mobile robotics.

\section*{Appendix A}

Differentiate Eq. \eqref{14} and insert Eq. \eqref{16}:
\begin{equation}
\small
\begin{aligned}
\label{18}
{\dot{V}_i} =& \frac{a_i}{\bar{a}_i} e_i u_{a,i}+\bar{a}^{-1}_i e_i F_i + \bar{a}^{-1}_i e_i D_i-\beta_i\hat{\Psi}^2_i-\xi_i \hat{\Psi}^2_i\\
&-|v_{c, i}|\hat{\Psi}^2_i-|v_i|\hat{\Psi}^2_i
\end{aligned}
\end{equation} 
Inserting Eq. \eqref{15}, knowing $e_i = v_{a,i} - v_{c,i}$ and $0<\frac{a_i}{\bar{a}_i}\leq 1$, and adding $-|v_{a,i}|+|v_{a,i}|$, we have:

\begin{equation}
\small
\begin{aligned}
\label{19}
{\dot{V}_i} \leq& -\lambda_i \frac{a_i}{\bar{a}_i} e_i (-|v_{a,i}|+|v_{a,i}|+ |v_i| - |v_{c,i}|) +\bar{a}^{-1}_i e_i F_i \\
&+ \bar{a}^{-1}_i e_i D_i-\beta_{i} \hat{\Psi}^2_{i}+ \frac{a_i}{\bar{a}_i}\hat{\Psi}^2_i|v_{a,i}-v_{c,i}|-\hat{\Psi}^2_i |v_{c,i}|\\
&- \hat{\Psi}^2_i |v_i|-\xi_i \hat{\Psi}^2_i
\end{aligned}
\end{equation} 
As $\frac{a_i}{\bar{a}_i}\leq 1$, $|v_{a,i}-v_{c,i}|\leq|v_{a,i}| + |v_{c,i}|$, $e_i = v_{a,i} -v_{c,i}$, and $|v_{c,i}|-|v_{a,i}|\leq |e_i|$: 
\begin{equation}
\small
\begin{aligned}
\label{20}
{\dot{V}_i} \leq & -\lambda_i \frac{a_i}{\bar{a}_i} e^2_i-\lambda_i \frac{a_i}{\bar{a}_i} e_i (-|v_{a,i}|+ |v_i|) +\bar{a}^{-1}_i e_i F_i + \bar{a}^{-1}_i e_i D_i\\
&-\beta_{i} \hat{\Psi}^2_{i}+\hat{\Psi}^2_i (|v_{a,i}|-|v_i|)-\xi_i \hat{\Psi}^2_i
\end{aligned}
\end{equation} 
Let us define: 
\begin{equation}
\small
\begin{aligned}
\label{21}
|\bar{a}^{-1}_i F_i|<F^*_i, \hspace{0.1cm} |\bar{a}^{-1}_i D_i|<D^*_i, \hspace{0.1cm} ||v_{a,i}|-|v_i||<<\xi_i
\end{aligned}
\end{equation} 
As the difference between the actual velocity $v_{a,i}$ and the measured signal $v_{i}$ is limited, we can define $\xi_i \in \mathbb{R}^+$ as large enough to satisfy $||v_{a,i}|-|v_i||<<\xi_i$. This assumption is significant because it allows control strategies to address valve faults without requiring infinite control effort. Thus:
\begin{equation}
\small
\begin{aligned}
\label{22}
{\dot{V}_i} \leq& - \lambda_{i} \frac{a_i}{\bar{a}_i} e^2_i+ |e_{i}|( F^*_i + D^*_i+\lambda_i \xi_i)-\beta_{i} \hat{\Psi}^2_{i} 
\end{aligned}
\end{equation} 
Using Young’s inequality \cite{shahna2024integrating, shahna2024robustqwewqde}:
\begin{equation}
\small
\begin{aligned}
\label{23}
\dot{V}_{i} \leq& - \lambda_{i} \frac{a_i}{\bar{a}_i} e^2_{i} + \frac{\sigma_i}{2} e_i^2 + \frac{(F^*_i + D^*_i+\lambda_i \xi_i)^2}{2 \sigma_i}  -\beta_{i} \hat{\Psi}^2_{i} 
\end{aligned}
\end{equation} 
where $\sigma_i \in \mathbb{R}^+$ is an arbitrary positive constant. Thus:
\begin{equation}
\small
\begin{aligned}
\label{24}
\dot{V}_{i} \leq& -(\lambda_{i}\frac{a_i}{\bar{a}_i}-\frac{\sigma_i}{2})   e^2_{i} + \frac{(F^*_i + D^*_i+\lambda_i \xi_i)^2}{2 \sigma_i}-\beta_{i} \hat{\Psi}^2_{i} 
\end{aligned}
\end{equation} 
where $\eta_{i} = \min [2\lambda_i a_i-\bar{a}_i\sigma_i,\hspace{0.1cm}  2 \beta_i]$. For any $\frac{2 \lambda_i }{\sigma_i}  > \frac{\bar{a}_i}{a_i}$, $\eta_i$ is always positive, and based on \eqref{14}, we have:
\begin{equation}
\small
\begin{aligned}
\label{25}
\dot{V}_i & \leq -\eta_i V_i + \frac{( F^*_i + D^*_i+\lambda_i \xi_i)^2}{2 \sigma_i}
\end{aligned}
\end{equation}
Now, we can extend the quadratic function \eqref{14} into all four wheels, as:
\begin{equation}
\small
\begin{aligned}
\label{26}
V & =\frac{1}{2}\sum_{i=1}^{4}\frac{1}{\bar{a}_i} \hspace{0.1cm} {e^2_{i}}+  \hat{\Psi}_{i}^2
\end{aligned}
\end{equation} 
Defining $\bm{a} = \text{diag}(\bar{a}^{-1}_1, \ldots, \bar{a}^{-1}_4):\mathbb{R} \rightarrow \mathbb{R}^{4\times 4}$, $\bm{e} = [e_1, \ldots, e_4]^{\top}:\mathbb{R} \rightarrow \mathbb{R}^4$, and $\bm{\hat{\Psi}} = [\hat{\Psi}_1, \ldots, \hat{\Psi}_4]:\mathbb{R} \rightarrow \mathbb{R}^4$:
\begin{equation}
\small
\begin{aligned}
\label{27}
V & =\frac{1}{2} \hspace{0.1cm} [\bm{e^\top} \hspace{0.05cm} \bm{a} \hspace{0.05cm} \bm{e}+ \bm{\hat{\Psi}^{\top}} \bm{\hat{\Psi}}]
\end{aligned}
\end{equation} 
From \eqref{25} and \eqref{26}, we have the derivative of \eqref{27}, as:
\begin{equation}
\small
\begin{aligned}
\label{28}
\dot{V} \leq& \sum_{i=1}^{4}- \eta_i V_i+ \frac{(F^*_i + D^*_i+\lambda_i \xi_i)^2}{2 \sigma_i}
\end{aligned}
\end{equation} 
Define $\eta = \min [\eta_{1}, \ldots, \eta_{4}] \in \mathbb{R}$. From \eqref{28}:
\begin{equation}
\small
\begin{aligned}
\label{29}
\dot{V} \leq& - \eta V+ \sigma
\end{aligned}
\end{equation} 
where $\sigma = \sum_{i=1}^{4}\frac{(F^*_i + D^*_i+\lambda_i \xi_i)^2}{2 \sigma_i}$.
Based on \cite{heydari2024robust}:
\begin{equation}
\small
\begin{aligned}
\label{30}
V \leq&  V\left(t_0\right) e^{-\left\{\eta\left({t-t_0}\right)\right\}}+{\sigma} \int_{t_0}^t e^{\left\{-\eta(t-T)\right\}}\hspace{0.2cm} dT
\end{aligned}
\end{equation}
Then:
\begin{equation}
\small
\begin{aligned}
\label{31}
V \leq&  V\left(t_0\right) e^{-\left\{\eta\left({t-t_0}\right)\right\}}+ \hspace{0.1cm}{\sigma} \hspace{0.1cm} {\eta}^{-1}
\end{aligned}
\end{equation}
Let us define the minimum and maximum eigenvalues of $\bm{a}$ as $\lambda_{\min }(\bm{a})$ and $\lambda_{\max}(\bm{a})$. Thus, $\lambda_{\min }(\bm{a})\|\bm{e}\|^2 \leq \bm{e^{\top}} \bm{a} \bm{e} \leq \lambda_{\max }(\bm{a})\|\bm{e}\|^2$. Hence, from \eqref{31}, we have:
\begin{equation}
\small
\begin{aligned}
\label{32}
\|\bm{e}\|^2 \leq & 2 (\lambda_{\min }(\bm{a}))^{-1} [V\left(t_0\right) e^{-\left\{\eta\left({t-t_0}\right)\right\}}+ \hspace{0.1cm}\sigma \hspace{0.1cm} {\eta}^{-1}]
\end{aligned}
\end{equation}
Based on Minkowski's inequality, we reach:
\begin{equation}
\small
\begin{aligned}
\label{33}
\|\bm{e}\| \leq \sqrt{
{2 (\lambda_{\min }(\bm{a}))^{-1} V\left(t_0\right)}} e^{-\frac{\eta}{2}(t-t_0)}+ \sqrt{{2 (\lambda_{\min }(\bm{a}))^{-1} \sigma \eta^{-1}}}
\end{aligned}
\end{equation}
This means that the Euclidean norm of the tracking linear velocity error vector exponentially converges to a stable region \cite{heydari2024robust, bahari2025system}. The radius of this region is given by $\sqrt{2\left(\lambda_{\min }(\boldsymbol{a})\right)^{-1} \sigma \eta^{-1}}$. This implies that the stability region will be larger if the intensity of faults, represented by $\sigma=\sum_{i=1}^4 \frac{\left(F_i^*+D_i^*+\lambda_i \xi_i\right)^2}{2 \sigma_i}$, is greater.

\bibliographystyle{IEEEtran}
\bibliography{main}

\end{document}